\documentclass{article}

\usepackage{blindtext}
\usepackage[a4paper, total={6in, 8in}]{geometry}
\usepackage[english]{babel}

\usepackage{amsmath}
\usepackage{makecell}
\usepackage{booktabs}
\usepackage{multirow}
\usepackage{float}
\usepackage{graphicx}
 \usepackage{cite}
\usepackage{algorithm}
\usepackage{color}

\providecommand{\keywords}[1]
{
  \small	
  \textbf{\textit{Keywords---}} #1
}

\begin{document}
\title{3D Super-Resolution Ultrasound with Adaptive Weight-Based Beamforming}
\author{Jipeng Yan, Bingxue Wang, Kai Riemer, Joseph Hansen-Shearer, \\
Marcelo Lerendegui, Matthieu Toulemonde, Christopher J Rowlands, \\
Peter D. Weinberg, Meng-Xing Tang,
\thanks{This work was supported by CZI project, EPSRC Impact Acceleration Account funding and MRC Confidence in Concept scheme at Imperial College, the Engineering and Physical Sciences Research Council (Grant No. EP/T008970/1).(\emph{Corresponding author: Tang.})}
\thanks{Yan, Wang, Riemer, Hansen-Shearer, Lerendegui, Toulemonde, and Tang are with Ultrasound Lab for Imaging and Sensing, Department of Bioengineering, Imperial College London, London, UK, SW7 2AZ. (e-mail: j.yan19@imperial.ac.uk;bingxue.wang18@imperial.ac.uk;  k.riemer16@imperial.ac.uk; joseph.hansen-shearer18@imperial.ac.uk; m.lerendegui@imperial.ac.uk; m.toulemonde@imperial.ac.uk; mengxing.tang@imperial.ac.uk).}
\thanks{Rowlands and Weinberg are with Department of Bioengineering, Imperial College London, London, UK, SW7 2AZ. (e-mail: c.rowlands@imperial.ac.uk; p.weinberg@imperial.ac.uk).}
}
\date{}
\maketitle
\begin{abstract}
Super-resolution ultrasound (SRUS) imaging through localising and tracking sparse microbubbles has been shown to reveal microvascular structure and flow beyond the wave diffraction limit. Most SRUS studies use standard delay and sum (DAS) beamforming, where large main lobe and significant side lobes make separation and localisation of densely distributed bubbles challenging, particularly in 3D due to the typically  small aperture of matrix array probes. This study aims to improve 3D SRUS by implementing a low-cost 3D coherence beamformer based on channel signal variance, as well as two other adaptive weight-based coherence beamformers: nonlinear beamforming with \emph{p}-th root compression and coherence factor.  The 3D coherence beamformers, together with DAS, are compared in computer simulation, on a microflow phantom, and in vivo.  Simulation results demonstrate that the adaptive weight-based beamformers can significantly narrow the main lobe and suppress the side lobes for modest computational cost. Significantly improved 3D SR images of microflow phantom and a rabbit kidney are obtained through the adaptive weight-based beamformers. The proposed variance-based beamformer performs best in simulations and experiments.
\end{abstract}
\keywords{Ultrasound localisation microscopy (ULM), super-resolution, contrast-enhanced ultrasound, 3D beamforming}

\section{Introduction}
Super-resolution (SR) ultrasound (US) through localisation and tracking of  microbubbles\cite{Couture2011MUSLI,Siepmann2011ImagingTumor}, also known as ultrasound localisation microscopy (ULM), is able to reconstruct images of the microvasculature with resolution beyond the wave diffraction limit in 2D \cite{viessmann2013acoustic,christensen2014vivo,errico2015ultrafast,ackermann2015detection,song2017improved,Paul20173DSR,zhu20193d,huang2021super,Provost2021vivo} and 3D \cite{christensen20173Dinvitro,foroozan2018_3ddecon,heiles2019ultrafast,JACCdemeulenaere2022coronary}. Compared to 2D, 3D imaging can offer volumetric coverage with super-resolution in all three dimensions, and does not have the out-of-plane motion that 2D imaging suffers. 
However, the image quality of full 3D US is generally not as good as in 2D. Matrix array probes used for generating 3D images typically require significantly more elements than 1D array probes, but the apertures of matrix arrays are commonly smaller due to the physical and economical constraints on the number of available channels in the acquisition equipment. As a consequence, the main lobe of the point spread function (PSF) in 3D US is usually wider than in 2D US. 
Row-Column array (RCA) probes have been designed for 3D US with larger apertures and fewer channels than matrix array probes\cite{demore2009firstRCAProbe,rasmussen20133DRCAProbe,jensen2019RCASR,joe2021ultrafast}, but the long element causes strong side lobes and distort the PSF in the nearfield/close to the probes.

\textcolor{black}{The acquisition time of SRUS based on localising sparse microbubbles heavily depends on the number of separatable bubbles within each image frames and desired level of information\cite{christensen2019poisson,hingot2019microvascular,dencks2020assessing}. Imaging with wide main lobes makes it difficult to isolate closely positioned bubbles, while strong side lobes  and noise can increase the risk of mistaking isolated patches for separated bubbles.} Exploring methods to reduce PSF main and side lobes and noise is consequently useful to reduce the acquisition time and increase the accuracy of localisation.

The sizes of the PSF main lobe and side lobes are affected not only by the transducer aperture size but also by beamforming algorithms. Delay and Sum (DAS) is the most common-used beamformer for high-frame-rate US, especially for 3D US \cite{provost20143dDAS,heiles2019ultrafast,JACCdemeulenaere2022coronary}. Apodisation can reduce PSF side lobes at the expense of an enlarged main lobe. Coherence-based beamformers have been demonstrated to reduce the PSF main lobe and side lobes and improve image quality over the DAS beamformer in 2D US \cite{matrone2019beamformingcomparison}.   There are different types of coherence-based beamformers. Some changes the method for combining channel signals into one pixel, such as the delay multiply and sum (DMAS) beamformer \cite{matrone2014DMAS} and short-lag spatial coherence beamformers \cite{lediju2011shortlag} which calculate pixel value by multiplication among channel signals before summation. Some adaptively changes the weights for each channel signals before summing them, such as the minimum variance (MV) \cite{synnevag2009MVbeamforming,diamantis2018MVforSRUS} and the nonlinear beamforming with \emph{p}-th root compression (\emph{p}-DAS) beamformers \cite{polichetti2018pDAS}. Some, such as the coherence factor (CF) beamformer\cite{li2003CFbeamforming}, apply adaptive weights to pixels. These advanced beamformers present significant opportunities to improve SRUS imaging in 3D, but also pose challenges due to the large number of channels acquired and millions of pixels to be reconstructed for one 3D US frame. 

Adaptive weight-based beamformers have only incremental computational costs compared to the standard  DAS beamformer. \textcolor{black}{ They are promising for 3D SRUS as their superiority over the DAS beamformer has been demonstrated on 2D US.} The CF beamformer is widely used and  
the \emph{p}-DAS beamformer has been proven to outperform other kinds of adaptive weight-based beamformer\cite{matrone2019beamformingcomparison}, such as phase coherence and sign coherence \cite{camacho2009phasecoherence}. 
Therefore, \emph{p}-DAS, CF and a computationally efficient 3D beamformer based on variance in the channel data are investigated in this study. They are evaluated through simulations and experiments.

\section{Methods}
In this section the different beamformers and their implementation are introduced, and this is followed by the description of the simulation and \emph{in vitro} and \emph{in vivo} experiments. All the coherence beamformers investigated in this study are based on the spatial coherence of channel signals from individual transducer elements. The beamfomers are implemented in CUDA (v10.0, Nvidia, CA, USA) and other data processing is done in MATLAB (2021a, MathWorks, MA, USA) on a desktop (CPU: AMD Ryzen 9 5900 Processor, GPU: Nvidia Geforce RTX3080, RAM: 128 Gb). 

\subsection{Delay and Sum}

Delay and Sum beamforming (DAS) reconstructs the image by

\begin{equation}
    y(t)=\sum_{n=1}^N s_n(t)
    \label{eq:DAS_noapo}
\end{equation}
where $s_n(t)$ is delay-cancelled signal for the \emph{n}-th channel; $N$ is the number of channels for the probe. To demonstrate the issues with 3D DAS, a matrix array probe (Vermon, Tours, France) with 32x32 elements, 7.8 MHz centra frequency and a 9.3 mm x 10.2 mm aperture imaging one point scatter is simulated in Field II\cite{jensen1996field}. A Hilbert transform is applied to each channel to obtain analytical signals. The delays of signals in each channel are then cancelled according to pixel locations \cite{montaldo2009coherent}, pulse length, and the transducer response in both transmit and receive, to centre the PSF at the scatter position. \textcolor{black}{ Namely, 1024 delay-cancelled channel signals generate one voxel value by beamforming.}

\begin{figure}
    \centering
   \includegraphics[width=8.5cm]{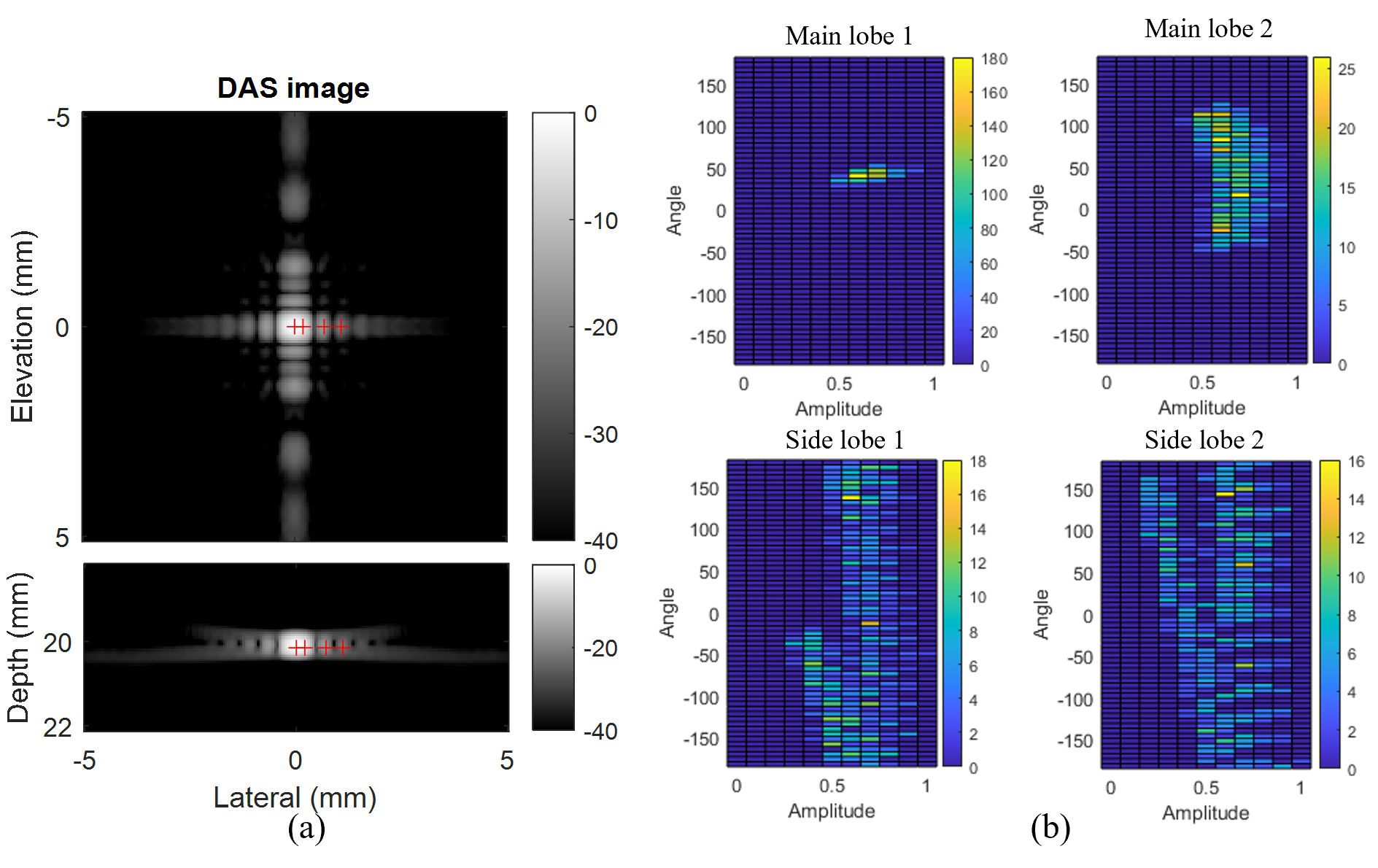}
    \caption{(a) B-mode image reconstructed by DAS. (b) Amplitude-angle histogram of delay-cancelled channel signals, where each pixel in the histogram is the number of channels with the same signal amplitude and phase angle. Four plots in (b) correspond to the four pixels labelled by red cross in (a). Note the colour bars have different ranges.}
    \label{fig:DAS_distribution}
\end{figure}

The brightness-mode (B-mode) 3D image of a point scatter reconstructed by DAS is shown in Fig. \ref{fig:DAS_distribution} (a). Four pixels are picked from the main lobe and side lobes, labelled by the red crosses, and denoted as main lobe 1-2, and side lobe 1-2 in order. The two-dimensional histogram of delay-cancelled signals corresponding to the four voxels is presented in Fig. \ref{fig:DAS_distribution} (b). Channel signals on the main lobe have similar amplitudes and phases, while channel signals on the side lobes have more dispersed amplitudes and phase angles. Therefore, summing up the channel signals would generate a main lobe brighter than the side lobes.

Element sensitivity apodisation $a_n(t)$, derived from the element directivity \cite{turnbull19912DApo}, is used for the DAS beamformer in the following comparison \textcolor{black}{to reduce the intensity of side lobes}. Namely, the implemented DAS beamformer is
\begin{equation}
    y(t)=\sum_{n=1}^N a_n(t)s_n(t)
    \label{eq:DAS_apo}
\end{equation}
 The sensitivity cut off is 0.5 for all the apodisation in this paper.

\subsection{Nonlinear Beamforming with p-th Root Compression}
A nonlinear beamformer based on  \emph{p}-th root compression (\emph{p}-DAS) \cite{polichetti2018pDAS} was designed by applying adaptive weights \begin{equation}
    w_n(t)=\frac{1}{|s_n(t)|^{\frac{p-1}{p}}}
    \label{eq:pDAS_weight}
\end{equation}
to channel signals followed by summation:
\begin{equation}
    \hat{y}_{pDAS}(t) =\sum_{n=1}^N w_n(t)s_n(t)
    =\sum_{n=1}^N sign(s_n(t))|s_n(t)|^{1/p}
    \label{eq:pDAS_root}
\end{equation}
where the sign function is used to maintain the polarity of signals.  Then, $\hat{y}_{pDAS}(t)$ is \emph{p}-powered to recover the amplitude of the signal
\begin{equation}
    y_{pDAS}(t) =sign(\hat{y}_{pDAS}(t))|\hat{y}_{pDAS}(t))|^{p}
    \label{eq:pDAS}
\end{equation}
As the \emph{p}-th root and \emph{p}-th power can distort the frequency components, band-pass filtering around the fundamental frequency is applied to the depth of $ y_{pDAS}(t)$ to remove the distorted components. The sampling frequency along depth needs to be sufficient to avoid aliasing of artificial harmonics. \textcolor{black}{In this study, $p$ is set to 4, with which $p$-DAS generated the best lateral resolution among all the adaptive weight-based beamformers in}  \cite{matrone2019beamformingcomparison}.

\subsection{Coherence Factor}
The Coherence factor (CF) is generally defined by the ratio of the coherent energy to the incoherent energy
\begin{equation}
    W_{CF}(t)=\frac{|\sum_{n=1}^N s_n(t)|^2}{\sum_{n=1}^N |s_n(t)|^2}.
    \label{eq:CF_weight}
\end{equation}

CF is calculated for each pixel and then used as an adaptive weight to the DAS image, i.e.,
\begin{equation}
 y_{CF} (t)=  W_{CF}(t)y(t)
    \label{eq:CF}
\end{equation}

The CF approaches one when channel signals are coherent with similar phases and it approaches zero if channel signals are random. For example, the histograms of the main lobe 1 and side lobe 1 in Fig.\ref{fig:DAS_distribution} (b) have a similar range of signal amplitude but a very different range of phase distribution. Applying CF weighting to the beamformer would therefore enhance the main lobe and reduce the side lobe. 

\subsection{Coherence Energy to Variance}
\begin{figure}
    \centering
    \includegraphics[width=7.5cm]{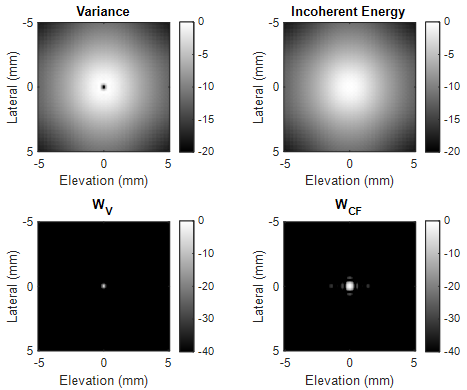}
    \caption{Demonstration of the principal difference between CF and CV beamformers. Images are plotted at the same lateral-elevation slice with the Fig. \ref{fig:DAS_distribution} (a). \textcolor{black}{ Each image is self-normalised and presented in the log-compressed scale. First row demonstrates the difference between the denominators of the W$_{V}$ and W$_{CF}$ weights. Second row demonstrates the difference between the two weights.}}
    \label{fig:diff_CF_CV}
\end{figure}

Distributions of channel signals, particularly the signal phase, become dispersed for pixels away from the scatter, as shown in Fig. \ref{fig:DAS_distribution}. In this study we explore using the variance, another measure of the coherence of the channel data, for beamforming. Variances in channel signals for each pixel of the image are shown as a variance map in Fig. \ref{fig:diff_CF_CV}.  An adaptive weight using this variance map is 

 \begin{equation}
    W_{VN}(t)=\frac{|\sum_{n=1}^N s_n(t)|^2}{\sum_{n=1}^N |{s_n(t)}-\frac{1}{N}\sum_{n=1}^N{s_n(t)}|^2}.
    \label{eq:V_weight_noapo}
\end{equation}
where the numerator is the coherent energy and denominator is the variance. To remove element sensitivity from $s_n(t)$ and recover the real variance in delay-cancelled signals, the denominator additionally incorporates inverse apodisation, where $s_n(t)$  is divided by the element sensitivity  $a_n(t)$. The above weight is rewritten as

\begin{equation}
    W_{V}(t)=\frac{|\sum_{n=1}^N s_n(t)|^2}{\sum_{n=1}^N |\frac{s_n(t)}{a_n(t)}-\frac{1}{N}\sum_{n=1}^N\frac{s_n(t)}{a_n(t)}|^2}.
    \label{eq:V_weight}
\end{equation}
The proposed weight is dimensionless. $W_V(t)$  can then be applied to the DAS beamformed voxel intensity $y(t)$
\begin{equation}
 y_{CV} (t)=  W_{V}(t)y(t)
    \label{eq:CV}
\end{equation}

A key difference between this Coherence to Variance (CV) beamformer and the CF beamformer is the subtraction of the mean channel signal in the denominator. As shown in Fig. \ref{fig:DAS_distribution}(b), distributions of channel signals in the centre and boundary of the main lobe (main lobe 1 and 2) are similar in amplitudes but different in phase angles. While the coherence factor reduces gradually when the pixel moves from the centre to the boundary of the PSF, variance can be close to zero at the centre of the PSF and hence creates a better defined main lobe peak and reduce side lobes in the CV weighting $W_{CF}$ as shown in Fig. \ref{fig:diff_CF_CV}. 

 \subsection{Simulation}
The four beamformers are compared in simulation.  The $W_{VN}$ weight is additionally investigated in the simulation to analyse the effectiveness of inverse apodisation; the corresponding beamformer is denoted as C$_{VN}$. The aforementioned  32×32 matrix array probe is simulated to image five scatters, using Field II. Gaussian windowed two-cycle sinusoidal waves are set as the transmission and reception response of each element of the probe. Two 1D Tukey windows with a factor of 0.5 are multiplied to generate a 2D apodisation for the transmission aperture. The matrix probe was set to transmit a single plane wave. The five scatters are placed on the central axis of the probe and spaced equally between 15 and 25 mm from the probe surface. The five scatters are given different scattering coefficients, decreasing from 1 to 0.2 with a step of 0.2.  The reconstructed voxel size  is 0.05×0.05×0.05 mm$^3$ cube.

Additive white Gaussian noise with an SNR of 0 dB and 10 dB are added to the radio frequency (RF) channel signals prior to image reconstruction respectively, to evaluate the Signal to Noise Ratio (SNR) of the five beamformers. After reconstruction, SNR is calculated by taking regions of noise only and comparing them to the averaged intensity of the centre PSF of the five targets. 
To quantitively compare the PSFs obtained by the five beamformers, Peak Side-to-Main Lobe Ratio (PSMR) and Full Width Half Maximum (FWHM) are calculated. Two kinds of PSMR are investigated. The conventional PSMR is defined by the peak ratio of the side lobe to the corresponding main lobe, and we named it as self-PSMR (SPSMR) for convenience. PSMR can also be calculated by the peak ratio of each side lobe to each different main lobe and is named cross-PSMR (CPSMR). The maximum of CPSMR is denoted as max-PSMR. In super-resolution imaging, \textcolor{black}{ a higher PSMR and a lower SNR give a higher risk of detecting isolated patches as single bubbles whereas a larger FWHM makes it more difficult to isolate close bubble images. The time taken to process single frame acquired with each beamformer is also listed for comparison.}

 \subsection{Experiment}
  \begin{figure}
    \centering
    \includegraphics[width=5.5cm]{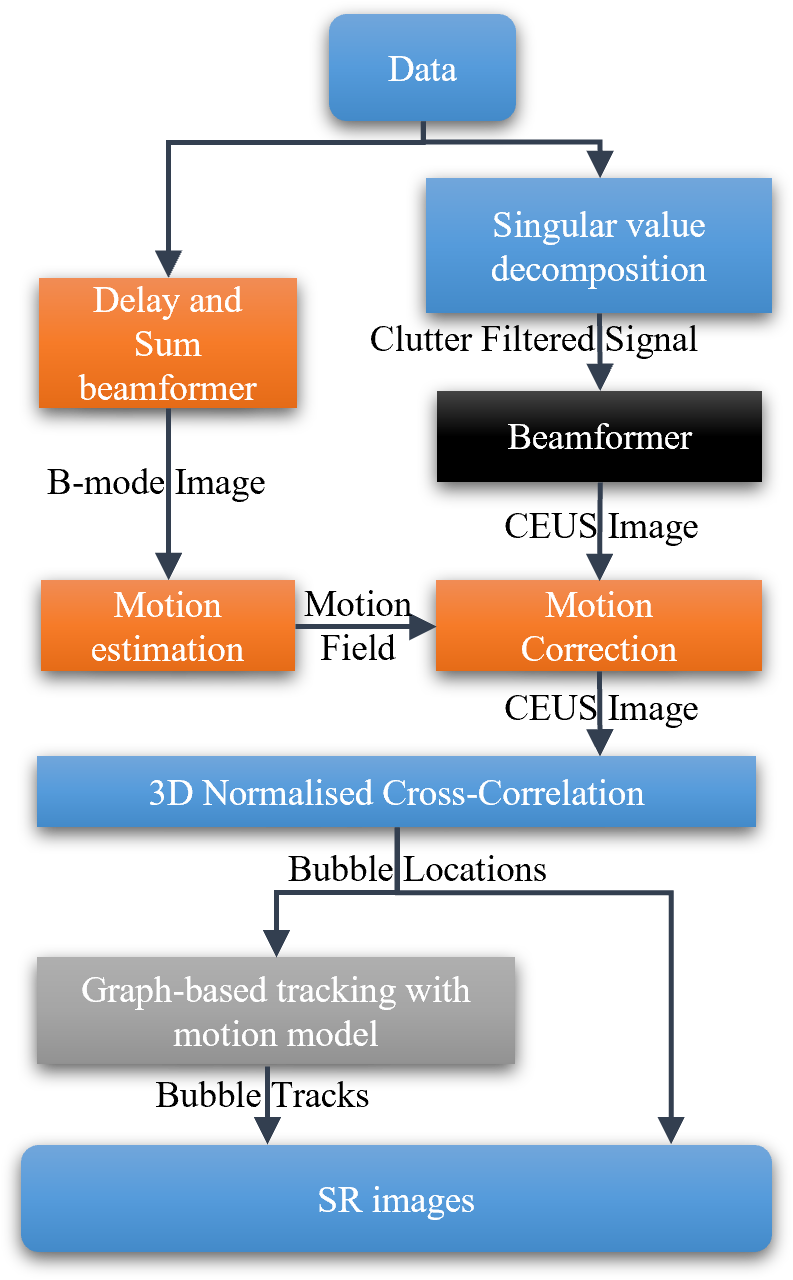}
    \caption{Data processing diagram. In vitro data go through the blue blocks and in vivo data additionally go through the orange blocks to correct tissue motions among frames. \textcolor{black}{Only the used beamfomer (black box) is different among the comparison.} Comparison is done without tracking to avoid the filtering effect of the tracking. A SR blood flow video of the rabbit kidney generated with the state-of-art bubble tracking algorithm \cite{yan2022super} is supplemented.}
    \label{fig:processingDiagram}
\end{figure}

 \subsubsection{In vitro phantom}
 An in-house prepared perfluorobutane microbubble solution \cite{li2015quantifying} was diluted to a concentration of $6\times 10^{6}$ bubbles/ml and pumped into two Hemophan cellulose tubes (Membrana, 3M, Wuppertal, Germany) with 200±15 µm inner diameter and 8±1 µm thickness wall using a syringe pump (Harvard Apparatus, Holliston, MA, USA). The two tubes were placed almost in parallel on the probe elevation-lateral projection plane and crossed with an angle of 3 degrees on the depth-lateral projection plane.
 
  \subsubsection{In vivo rabbit kidney}
  A specific-pathogen-free New Zealand White rabbit (male, HSDIF strain, age 13 weeks, weight 2.4 kg, Envigo, UK) was sedated with acepromazine (0.5 mg/kg, i.m.) and anaesthetized with medetomidine (Domitor, 0.25 mL/kg, i.m.) plus ketamine (Narketan, 0.15 mL/kg, i.m.). Anaesthesia was maintained for approximately 4 hours by administration of 1/3 of the initial medetomidine and ketamine dose every 30 minutes.  Bubbles were injected with a first bolus of 0.1 ml and following boluses of 0.05 ml.
  
  To access the renal vasculature, the rabbits were shaved and positioned supine. Following tracheotomy, mechanical ventilation was given at 40 breaths/minute. Body temperature was maintained with a heated mat. Oxygenation and heart rate were continuously monitored. The experiment complied with the Animals (Scientific Procedures) Act 1986 and was approved by the Animal Welfare and Ethical Review Body of Imperial College London.
  
  \subsubsection{Imaging sequence}
  Data of the phantom and rabbit kidney are  acquired with the Vantage 256 ultrasound research system (Verasonics Inc., Kirkland, WA, USA) and with the matrix array probe. As there are 1024 elements in the probe and only 256 channels in the research system, a multiplexing technique is used, in which 1024 elements are excited at the same time to transmit plane waves and four sub-apertures, i.e., 256 elements in each, receive signals in sequence for one frame. Images are acquired at a frame rate of 500 Hz without angle compounding. 1000 frames of \emph{in vitro} data and 1500 frames of \emph{in vivo} data were acquired. 
  
\subsubsection{SR Processing pipeline}
Data processing is conducted according to the pipeline shown in Fig. \ref{fig:processingDiagram}.
Tube and tissue background signals are removed by applying singular value decomposition (SVD) to channel signals and  reconstructed channel signals are then beamformed by the four techniques to obtain the contrast enhanced ultrasound (CEUS) sequence.  Voxels of images reconstructed by beamformers have dimensions of 0.05×0.05×0.05 mm$^3$ cube for \emph{in vitro} experiment and 0.1×0.1×0.1 mm$^3$ cube for \emph{in vivo} experiment to fit the data to available memory. Image intensities are normalised by the maximum of each beamformed sequence. 
    Thresholds for background noise removal are adjusted for DAS, \emph{p}-DAS, CF and, CV images to achieve similar numbers of localised events for SR images.
    \textcolor{black}{ Then, comparison is done on the SR density map, based on the assumption that with a similar amount of localisations, a better beamformer can provide fewer wrong/false localisations and potentially reconstruct more tube/micro-vessel structures.}

 The PSFs for each beamformer are estimated by simulations. 3D normalised cross-correlation (3DNCC) is then implemented by convolving the the background-noise-removed images with flipped PSFs.
Noise is further removed by thresholding the 3DNCC coefficient maps at 0.3.  Peaks on the 3DNCC coefficient maps are detected by the MATLAB ‘imregionalmax’ function and the coefficient maps are cropped by a 5 x 5 x 5 window around the peaks into small patches. Super-resolution localisation is performed by detection of the maximum in each small patch after interpolating each patch using a cubic spline and voxels 1/10 of the original size. All localised bubbles are accumulated to generate the SR bubble density map. For \emph{in vivo} data processing, a two-stage image registration algorithm \cite{harput2018two} is used to correct motions, which was induced mainly by animal breathing. Motion is estimated on B-mode images reconstructed by the DAS beamformer and the resulting motion fields are applied to the CEUS images reconstructed by the four beamformers.

\begin{figure*}
    \centering
    \includegraphics[width=16cm]{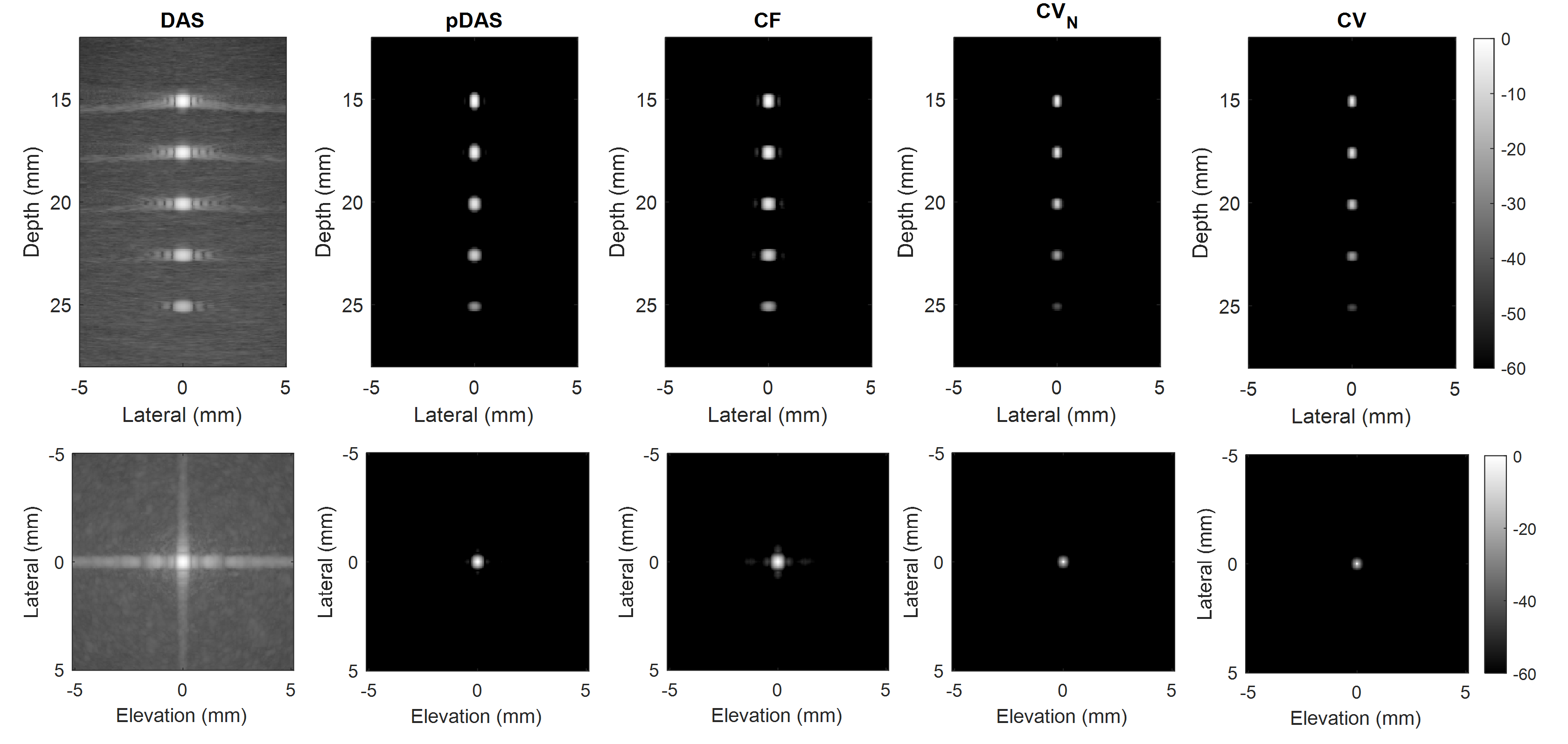}
    \caption{Maximum intensity projection (MIP) of 3D images reconstructed by the five different beamformers  \textcolor{black}{with channel SNR of 0 dB}.} 
    \label{fig:simulated_image}
\end{figure*}

\begin{table*}[h]
\caption{Quantified metrics for five beamformers at 0 dB SNR in channels.}
\centering
\begin{tabular}{cccccc}
\toprule
 &DAS &\emph{p}-DAS &CF &CV$_N$ &CV\\
\midrule

SPSMR(dB) &-16.2±1.51 &-53.2±1.30 &-42.3±2.44 &-60.7±10.2 &-61.9±11.3\\

Max-PSMR(dB) &-2.26 &-27.2 &-22.2 &-29.7 &-32.4\\

Lateral FWHM(mm) &0.556±0.084 &0.279± 0.052 &0.326±0.054 &0.176±0.095 &0.166±0.098\\

Elevation FWHM(mm) &0.510±0.077 &0.251±0.047 &0.296±0.049 &0.159±0.086 &0.151±0.089\\

SNR(dB) &27.4 &101.0 &74.1 &84.7 &86.6\\
Processing Time(s) &0.68 &4.83 &0.82 &1.49 &1.49 \\
\bottomrule
\end{tabular}
  \label{tb:simulation_1}
\end{table*}

\begin{table*}[h]
\caption{Quantified metrics for five beamformers at 10 dB SNR in channels.}
\centering
\begin{tabular}{cccccc}
\toprule
 &DAS &\emph{p}-DAS &CF &CV$_N$ &CV\\
\midrule

SPSMR(dB) &-16.0±1.25 &-52.6± 0.39 &-42.7±1.61 &-69.7±5.24 &-73.0±7.32\\

Max-PSMR(dB) &-1.95 &-34.6 &-27.3 &-47.4 &-50.3\\

Lateral FWHM(mm) &0.558±0.088 &0.278± 0.051 &0.328±0.056 &0.107±0.048 &0.095±0.051\\

Elevation FWHM(mm) &0.507±0.078 &0.250±0.046 &0.296±0.050 &0.097±0.044 &0.088±0.046\\

SNR(dB) &36.9 &113.0 &84.0 &97.2 &100.0 \\
Processing Time(s) &0.69 &4.86 &0.81  &1.50 &1.51\\
\bottomrule
\end{tabular}
  \label{tb:simulation_2}
\end{table*}

\section{Results}
\subsection{Simulation}
\begin{figure}
    \centering
    \includegraphics[width=9cm]{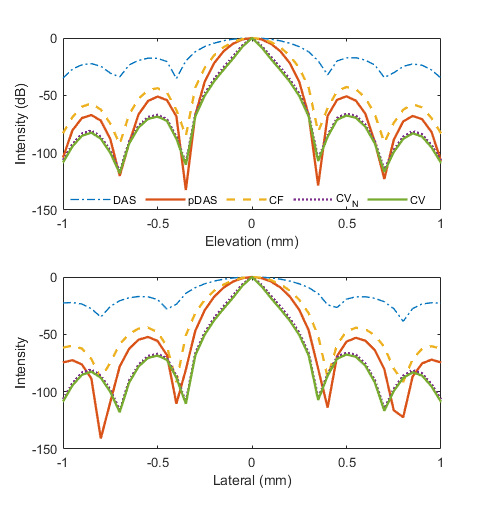}
    \caption{Elevational and lateral profiles of PSF reconstructed by different beamformers for the scatter located at 17.5 mm.}
    \label{fig:simulated_profile}
\end{figure}

Maximum intensity projections (MIPs) of the 3D images reconstructed with channels with SNR of 0 dB by the five beamformers are shown in Fig. \ref{fig:simulated_image}.  The image obtained by the CV beamformer is visually the best with narrow main lobes and small side lobes. The lateral and elevational profiles are shown in Fig. \ref{fig:simulated_profile}.

Quantitative metrics for the five beamformers are given in Table \ref{tb:simulation_1} and \ref{tb:simulation_2}, where the mean and standard deviations of SPSMRs and FWHMs are listed for the five scatters. Note that the max-PSMR is much higher than the SPSMR which means side lobes of strong-reflection scatters can be mistaken for single bubbles when strong- and weak- scatters appear in the same frame. While the max-PSMR is around -2.0 dB for conventional DAS, adaptive weight-based beamformers can achieve at least a 10-fold improvement in max-PSMR at both noise levels. Additionally, compared to lateral and elevational FWHMs obtained by DAS, CF’s have an approximately 40\% reduction in the FWHMs; \emph{p}-DAS’s achieve around a 50\% reduction and CV’s reach an 80\% reduction with channel SNR of  10 dB. SNRs in images are significantly improved by the five coherence-based beamformers due to the large number of channels and SNRs of the adaptive weight-based beamformers can obtain at least a 45 dB improvement over conventional DAS. In terms of computational cost, \emph{p}-DAS, CF and the variance-based beamformers require 6.9, 1.2 and 2.2 times the resource of the DAS beamformer respectively.

\begin{figure}
    \centering
    \includegraphics[width=8cm]{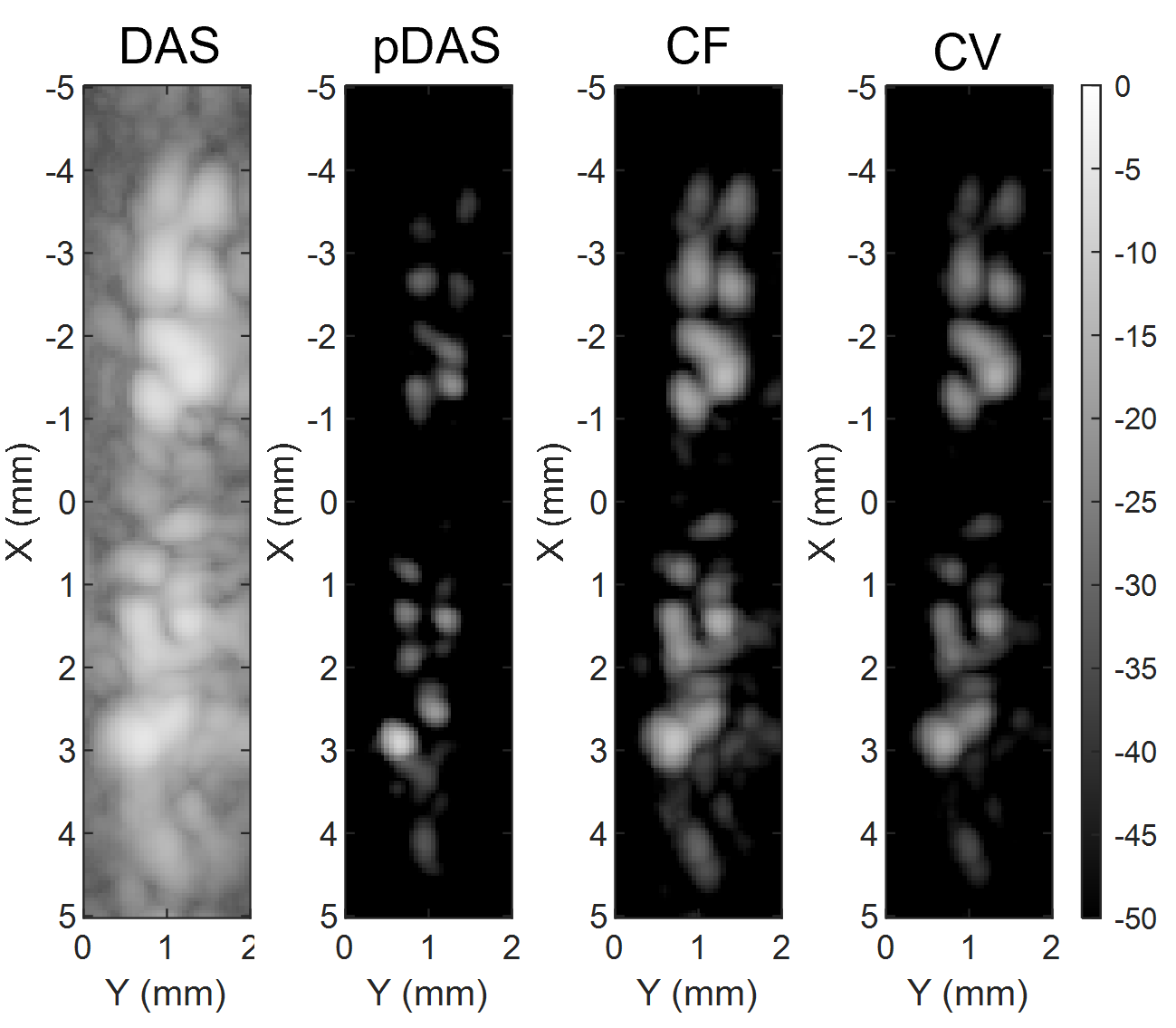}
    \caption{Lateral-elevation MIPs  of bubble images beamformed by DAS, \emph{p}-DAS, CF, and CV.}
    \label{fig:invitro_bubbleimage}
\end{figure}

\begin{figure}
    \centering
    \includegraphics[width=7cm]{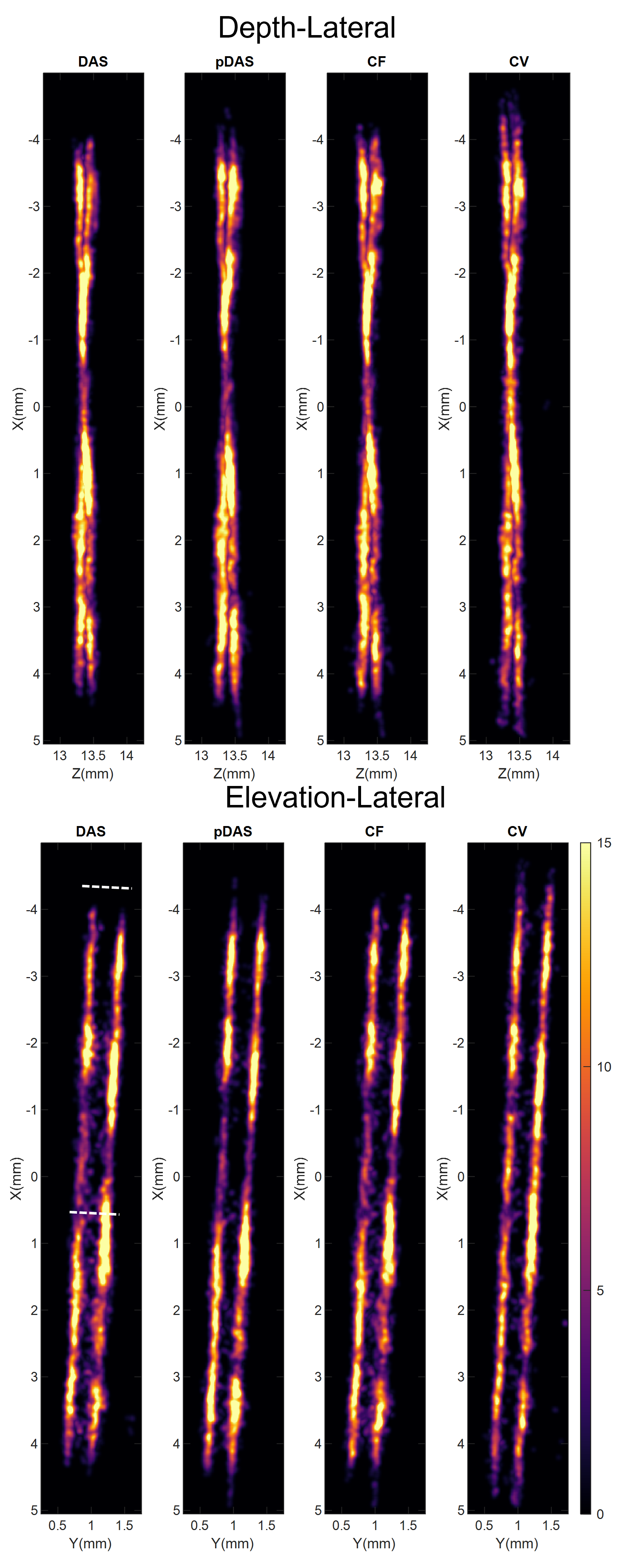}
    \caption{MIPs of SRUS density images obtained accumulating  about 6500 bubbles detected on images reconstructed by the four beamformers.}
    \label{fig:invitro_srimage}
\end{figure}

\subsection{In vitro experiment}

Bubble images reconstructed from clutter-filtered signals by the four beamformers are shown in Fig. \ref{fig:invitro_bubbleimage}. DAS gives the worst image quality. 6439, 6413, 6829 and 6469 bubbles can be localised by using thresholds of -10 dB, -35 dB, -27.5 dB, and -40 dB for the DAS, \emph{p}-DAS, CF, and CV images respectively. SRUS images obtained by accumulating these bubbles are shown in Fig. \ref{fig:invitro_bubbleimage}.  There are fewer wrong localisations between the two tubes for the adaptive weight-based beamformers than for the DAS beamformer. The SRUS image obtained with the CV beamformer gives the highest signal at around 0 mm in the lateral direction where the two tubes are the closest.

\begin{figure}
    \centering
    \includegraphics[width=8.5cm]{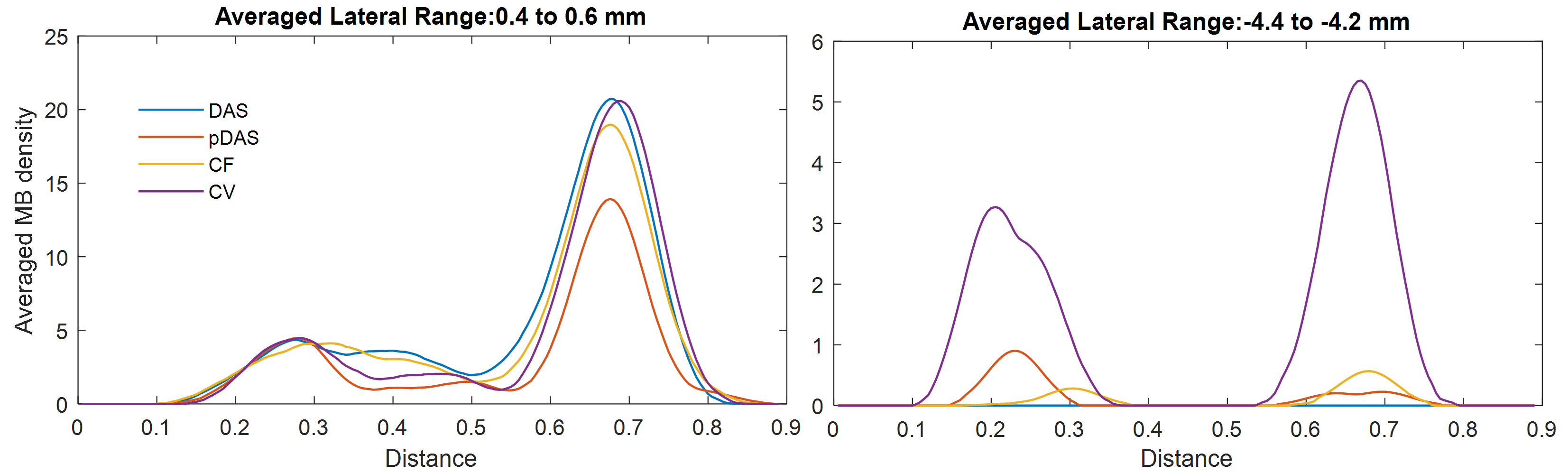}
    \caption{MB density profile averaged across the tube direction on the MIP image shown in Fig.\ref{fig:invitro_srimage}}
    \label{fig:invitro_srprofile}
\end{figure}

Microbubble density profile across the two regions, labelled by white lines in the CV image, are shown in Fig. \ref{fig:invitro_srprofile}.  For the region from 0.4 mm to 0.6 mm, the two tubes cannot be clearly separated in profiles of the SRUS image using the DAS beamformer. In contrast, the two peaks are distinguishable in the SRUS image profiles obtained using the adaptive weight-based beamformers. For the region from -4.4 to -4.2 mm in the $x$ direction, the DAS SRUS image completely loses the tube signal, and the profile of the CV SRUS presents the highest two peaks.

\subsection{In vivo experiment}
\begin{figure*}
    \centering
    \includegraphics[width=13cm]{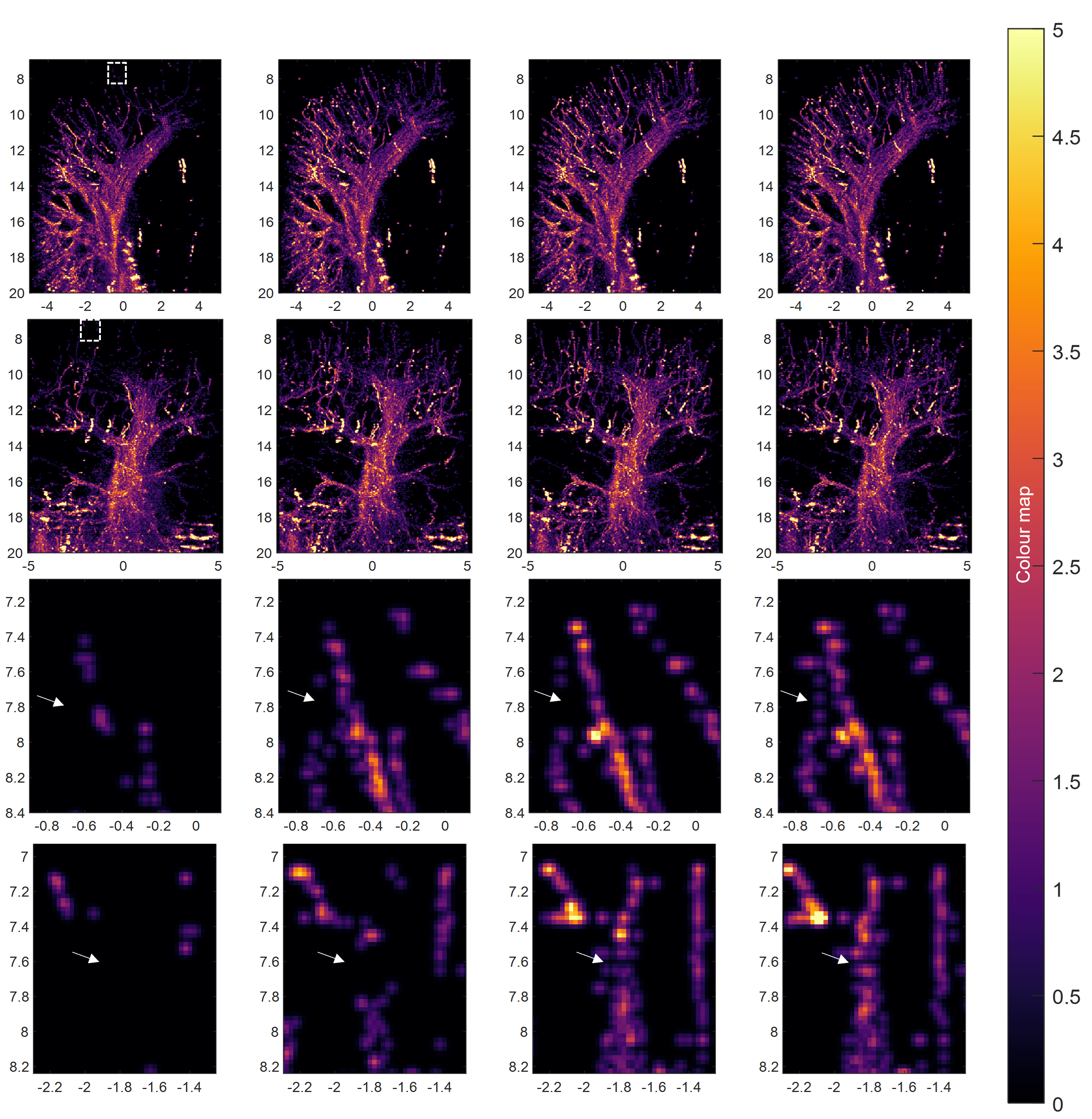}
    \caption{MIPs of SRUS images of rabbit kidney. The four columns correspond to the DAS, \emph{p}-DAS, CF, and CV beamformers, from left to right. First row: lateral-depth-plane MIPs; Second row: elevation-depth-plane MIPs; Third and fourth rows: Zoomed-in images of the area in the boxes shown in the first and second rows respectively.}
    \label{fig:invivo_srimage}
\end{figure*}

 74842, 67495, 69292 and 63660 bubbles are localised by using thresholds of -14 dB, -42.5 dB, -32.5 dB, and -35 dB for DAS, \emph{p}-DAS, CF, and CV images respectively. 
 MIPs of 3D in vivo SRUS images are shown in Fig. \ref{fig:invivo_srimage}. \textcolor{black}{ More dense vessels in the top region and less noise in the bottom region can be detected with the adaptive weight-based beamformers, compared to those with the conventional DAS.} 
 In the magnified images in the third and fourth rows, all the adaptive weight-based beamformers provides more SRUS vascular signals than the DAS beamformer. Also, the CV beamformer generates some additional localisations that are missed by the other methods.

\section{Discussion}
In this study advanced beamformers for 3D SRUS imaging are investigated. The results
show that the adaptive weight-based beamformers can significantly narrow the PSF main lobe and suppress the side lobes and noise, and are able to improve 3D SRUS images \emph{in vitro} and \emph{in vivo} with modest computational cost. They also show that the proposed variance-based beamformer performs best both in simulations and experiments.  Compared to CV$_{N}$ \cite{wang2018MTSD}, CV generates better image resolution and contrast measures in simulation by considering channel element sensitivity.

Improved separation and localisation of individual bubbles at high bubble concentration reduce acquisition time and hence the impact of motion on SRUS imaging. Compared to most existing SRUS studies that improved isolation or localisation algorithms, such as deconvolution \cite{bar2018sushi,foroozan2018_3ddecon,solomon2019exploiting,qian2020vivo,yan2022super} and normalised cross-correlation \cite{heiles2019ultrafast,JACCdemeulenaere2022coronary,song2017improved}, this study aims to improve 3D SRUS through advanced beamforming, which also has the potential to be combined with the existing algorithms. Classical DAS beamforming has limited image resolution and contrast. Consequently individual bubble signals can be difficult to separate at high bubble concentration. Furthermore, the side lobes of a bright bubble can be similar in intensity to a dim bubble, potentially leading to false localisations.  While the main lobe can be narrowed by coherent compounding and side lobes compressed by apodisation, the improvement is at the expense of frame rate or resolution. Additionally, the small aperture of a typical matrix array limits improvements when the steering angle is large.

Adaptive weight-based beamformers are investigated for improving, because of their relatively low computation cost which is valuable for dealing with the large data size from matrix array probes. For example, \emph{p}-DAS beamformer calculates the adaptive weight for each channel by using the single channel signal; the CF beamformer calculates the adaptive weight for each pixel across a single combination of the channels.  It is worth noting that images in the \emph{p}-DAS beamformer had to be reconstructed with a voxel of 0.01 mm  depth to mitigate aliasing of the band-pass filtering and then down sampled to the same voxel size with the other beamformers. Denser voxels make the \emph{p}-DAS beamformer take much more time than the other two adaptive weight-based beamformers. In contrast, DMAS \cite{matrone2014DMAS} and short-lag spatial coherence beamformers \cite{lediju2011shortlag} need to multiply signals between different combinations of channels and the number of combinations dramatically increases with channel numbers.

It should be noted that all the methods use the 3D normalised cross-correlation with the corresponding simulation PSF to localise 3D bubbles \cite{heiles2019ultrafast,JACCdemeulenaere2022coronary}. Simulated PSFs do not consider some of the physics, such as the nonlinearity of bubbles and phase aberrations, which may explain difference between simulation and experiment. As a result, the cross-correlation between the estimated PSF and the experimental data is low and thus 0.3 is selected as the cross-correlation coefficient threshold to include enough signals for the SRUS reconstruction.

For the SRUS image of the cross-tube phantom there should be no bubbles outside of the two tubes but the area between the tubes is not totally black due to false separations and localisations (Fig. \ref{fig:invitro_srimage}).
SRUS images derived from the adaptive weight-based beamformers have fewer wrong localisations than the baseline.  The intensity profiles also demonstrate the benefit of using adaptive weight-based beamformers for 3D SRUS imaging to obtain more distinguishable structures and the superior performance of the CV beamformer, which gives the longest presented tube images and the most localisations in  vessels close to the probe. 

\section{Conclusion}
This study compares four coherence-based beamformers for 2D US matrix array probes to improve 3D SRUS imaging. Simulations and experiments verify the superiority  of the adaptive weight-based beamformers over the DAS beamformer for single-plane-wave transmission. The proposed CV beamformer gives the narrowest main lobes and weakest side lobes in simulations and generate the best \emph{in vitro}  and \emph{in vivo} SRUS images.

\section{ACKNOWLEDGEMENTS}
We would like  to thank Prof. Chris Dunsby from Imperial College London for  insightful discussion.

\bibliographystyle{IEEEtran}
\bibliography{sample}

\end{document}